\documentclass[reprint, twocolumn, nofootinbib]{revtex4-1}

\usepackage{graphicx}
\usepackage{dcolumn}
\usepackage{bm}
\usepackage{tabularx} 
\usepackage{amsmath}
\usepackage{caption}
\usepackage[title]{appendix}

\usepackage{tikz}
\usepackage{pgfplots}
\pgfplotsset{width=250pt,compat=1.9}
\usepgfplotslibrary{fillbetween}

\usepackage{listings}
\usepackage{units}
\usepackage{numprint}

\usepackage{balance}

\begin{document}

\title{Sensitivity potential to a light flavor-changing scalar boson with DUNE and NA64$\mu$ \\}
\vspace{\baselineskip}
\author{B. Radics}
\email[Correspondence email address: ]{bradics@yorku.ca}
\affiliation{
Department of Physics and Astronomy, York University, Toronto, ON, Canada
}%
\author{L. Molina-Bueno}
\affiliation{
Instituto de F\'isica Corpuscular, Universidad de Valencia and CSIC, Carrer del Catedtr\'atic Jos\'e Beltr\'an Martinez, 2, 46980 Paterna, Valencia, Spain 
}
\author{L. Fields}
\affiliation{Department of Physics and Astronomy, University of Notre Dame, IN, 46556, USA}
\author{H. Sieber}
\author{P. Crivelli}
\affiliation{ETH Z\"urich, Institute for Particle Physics and Astrophysics, CH-8093 Z\"urich, Switzerland}

\begin{abstract}
  In this work, we report on the sensitivity potential of complementary muon-on-target experiments to new physics using a scalar boson benchmark model associated with charged lepton flavor violation. The NA64$\mu$ experiment at CERN uses a 160-GeV energy muon beam with an active target to search for excess events with missing energy and momentum as a probe of new physics. At the same time, the proton beam at Fermilab, which is used to produce the neutrino beam for the Deep Underground Neutrino Experiment (DUNE) will also produce a high-intensity muon beam dumped in an absorber. Combined with the liquid Argon Near Detector, the system could be used to search for similar scalar boson particles with a lower energy but higher intensity beam. We find that both NA64$\mu$ and DUNE could cover new, unexplored parts of the parameter space of the same benchmark model, providing a complementary way to search for new physics.

\end{abstract}
\maketitle


\section{\label{sec:Intro}Introduction}

Observations in cosmology and astrophysics imply the existence of a Dark Sector potentially containing new particles that could weakly couple to Standard Model (SM) particles \cite{doi:10.1146/annurev-nucl-102419-055056}. Neutrino oscillations coupled with non-zero neutrino masses provide an experimental evidence of lepton-flavor violation. Furthermore, the existing discrepancy between the measured \cite{PhysRevLett.126.141801} and expected \cite{AOYAMA20201,BorsanyiNATURE} value of the muon anomalous magnetic moment provides a strong motivation for new physics searches with muons \cite{PhysRevD.95.115005}. Inspired by these developments, a certain class of new theories proposes the search for charged lepton flavor violation (CLFV), which is heavily suppressed in the Standard Model (SM). In the coming decades, a new generation of experiments will conduct experimental searches for CLFV \cite{Meucci:20229n, universe7120466, Mu2e:2014fns, Hesketh:2022wgw, COMET:2018auw, Belle-II:2018jsg}. In parallel, next-generation long-baseline neutrino oscillation experiments \cite{PhysRevD.105.072006, 2015PTEP.2015e3C02A} will study neutrino oscillations and at the same time produce an intense muon beam-dump with its beam line. 

In this work we study the sensitivity potential of muon-on-target experiments to new physics using a CLFV benchmark model. The physics scenario was introduced in a recent work \cite{Pospelov23} and uses a light, scalar boson associated with $\mu-\tau$ conversion. While the previous work derived constraints from data in existing beam-dump experiments (LSND\cite{PhysRevD.63.112001}, NuTeV \cite{PhysRevLett.87.041801}, CHARM \cite{DORENBOSCH1986473}) and for the future SHiP experiment \cite{Alekhin_2016}, here we focus on two further experiments: the coming neutrino experiment, DUNE \cite{PhysRevD.105.072006} deploying a high-intensity beam from the Long-Baseline Neutrino Facility (LBNF) \cite{abi2020deep}, and a fixed-target experiment, NA64$\mu$ \cite{Gninenko:2653581,GninenkoNA64mu}, which searches for new physics with the high-energy muon beam from the CERN Super Proton Synchrotron (SPS) accelerator. Hence, we study new physics searches using two complementary muon-beam setups.  Even though we focus on one particular benchmark model scenario, there are also other possibilities to probe hidden sectors with muon beams and using similar techniques \cite{ Gninenko:2018ter,Gninenko:2018num,Gninenko:2019qiv,Gninenko:2022ttd,sieber2023probing}.

\section{\label{sec:Model}Flavour-changing scalar with long lifetime}

The model proposed by \cite{Pospelov23} considers a new complex scalar field, $\phi$, with a mass window $[m_{\tau} - m_{\mu}, m_{\tau} + m_{\mu}]$ that couples to $(\mu, \tau)$. With such a mass range, long lifetimes can be achieved with propagation distances on the order of tens of kilometers. The effective Lagrangian interaction terms describe the coupling of the new scalar field with leptons,
\begin{equation}
\mathcal{L_{I}} = \phi \bar{\mu}(g_{V} + g_{A}\gamma^{5})l + \phi^{*}\bar{l}(g_{V}^{*} - g_{A}^{*}\gamma^{5})\mu
\end{equation}
with vector and axial vector coupling $g_{V}$, $g_{A}$. This model produces a benchmark parameter region explaining the muon $g-2$ anomaly with a typical value of $|g_{V}|\simeq 3\times 10^{-3}$, also used in this work. There are multiple production modes to produce $\phi$ at beam dump experiments: direct electroweak process, heavy meson decay and high-energy muons on hitting a fixed target. In this work, we focus on the third case (the so-called $\mu$-on-target scenario). In this mode, muons pass through dense material and could produce the $\phi$ boson via the exchange of a virtual photon with the nuclei of the target, $\mu(p) N(P_i) \rightarrow \tau(p') \phi(k) N(P_f)$. Once created, the bosons produce a missing energy signature or propagate long distances and may be detected in a detector downstream from the target. When the incoming beam energy is much higher than the particle mass, the double-differential cross-section of the $2\rightarrow 3$ production process can be estimated using the equivalent photon approximation \cite{PhysRevD.80.075018},
\begin{align}\label{Eq:EPAequation}
\frac{d^{2}\sigma(p+P_{i}\rightarrow p' + k + P_{f})}{dE_{\phi}d\cos\theta_{\phi}}&=\\ \nonumber
\frac{\alpha\chi}{\pi}\frac{E_{\mu}x\beta_{\phi}}{1-x}&\frac{d\sigma(p+q\rightarrow p' + k)}{d(p\cdot k)}\biggr\rvert_{t=t_{\mathrm{min}}}
\end{align}
where $E_{\mu}$ is the initial muon energy, $E_\phi$ is the energy and $\theta_{\phi}$ the scattering angle of $\phi$ with respect to the initial muon in the lab frame, $x = E_{\phi}/E_\mu$, $q = P_i - P_f$, $t = -q^2$ is the momentum transfer, $\alpha = 1/137$ is the fine structure constant, $\beta_{\phi}$ is the relativistic factor for $\phi$, and $\chi$ is the effective photon flux, defined as 
\begin{equation}
    \chi = \int_{t_{\mathrm{min}}}^{t_{\mathrm{max}}} dt\frac{t - t_{\mathrm{min}}}{t^2} F^2(t)
\end{equation}
where $F(t) = Z^2/(1+t/d)^2$ is the form factor with $d = 0.164$ GeV$^2 A^{-2/3}$, and the limits are given in the appendix of \cite{Pospelov23}. We calculate the analytical expression for $\chi$ using Mathematica \cite{Mathematica} and obtain 
\begin{equation}
    \chi = Z^2 \left[\frac{t_{\mathrm{min}}}{t} + \frac{d+t_{\mathrm{min}}}{d+t} + \frac{(d+2t_{\mathrm{min}})}{d}\frac{\ln{t}}{\ln{(d + t)}} \right]_{t_{\mathrm{min}}}^{t_{\mathrm{max}}}.
\end{equation}

In Eq.~\ref{Eq:EPAequation}, $\sigma(p+q\rightarrow p' + k)$ is the cross section for the $2\rightarrow 2$ scattering process, $\mu(p)\gamma(q)\rightarrow \tau(p')\phi(k)$,
\begin{equation}
\frac{d\sigma(p+q\rightarrow p'+k)}{d(p\cdot k)} = \frac{|\mathcal{\bar{A}}_{2\rightarrow 2}|^2}{8\pi s^2}
\end{equation}
where $|\mathcal{\bar{A}}|^2$ is the amplitude squared, which we calculate using the FeynCalc tools \cite{SHTABOVENKO2020107478} for Mathematica. We obtain the following expression for the $(\mu, \tau)$ case,
\begin{align} 
&|\mathcal{\bar{A}}_{2\rightarrow 2}|^2  = -\frac{e^{2} m_\mu m_\tau (g_{A} g_{A}^* - g_{V} g_{V}^*)}{(m_\mu^2 - s)^2 (m_\tau^2 - u)^2} \times \\ \nonumber
& \times [ m_\mu^4 (m_\phi^2 + u)+ 2 m_\mu^3 (m_\tau^3- m_\tau u)\\ \nonumber
&+ m_\mu^2 \left( m_\tau^4 - 2 m_\phi^2 s - 2m_\tau^2 u + u(u-2s) \right) \\ \nonumber
& + 2 m_\mu m_\tau s(u - m_\tau^2 ) + s \left( m_\phi^2 s + m_\tau^4 - 2 m_\tau^2 u + u(s+u) \right) ] \\ \nonumber
\end{align}
where $e = \sqrt{4\pi\alpha}$, $m_{\phi}$, $m_{\tau}$, $m_\mu$ are the masses of the boson $\phi$, $\tau$ and $\mu$, respectively, and $s$, $u$ are the Mandelstam variables, which can be evaluated in the laboratory frame,
\begin{align}\label{Eq:Mandelstam}
s = (p+q)^2 \simeq m_\mu^2 - \frac{u - m_{\tau}^2}{1 - x} \\ \nonumber
u = (p-k)^2 \simeq -E_\mu x \theta_\phi^2 - \frac{1-x}{x} m_{\phi}^2 + (1-x)m_\mu^2.
\end{align}

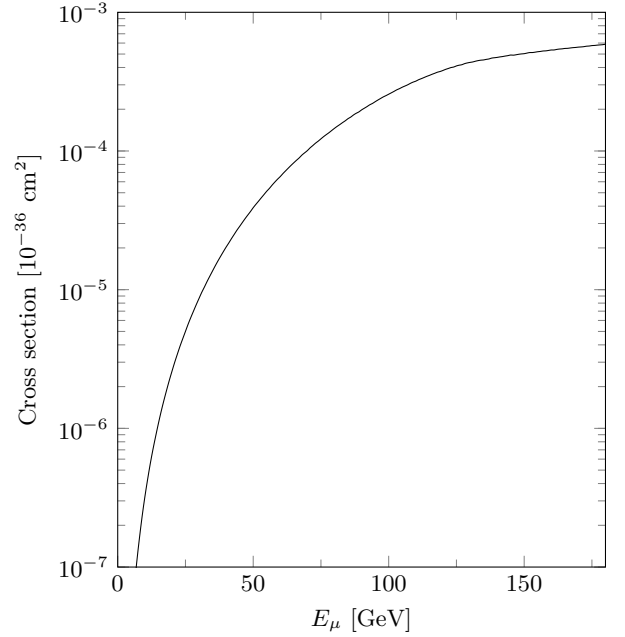
\begin{figure}
\begin{tikzpicture}
\begin{semilogyaxis}[
    xlabel={$E_{\mu}$ [GeV]},
    ylabel={Cross section [$10^{-36}$ cm$^2$]},
    xmin=0, xmax=180,
    ymin=1e-07, ymax=0.001,
    minor tick num = 1,
    width = 0.45\textwidth,   
    height = 0.5\textwidth,
    legend pos=north east,
    ymajorgrids=false,
    grid style=dashed,
]

\addplot [
    smooth,
    thin,
    color=black
    ] table {Figs/CrossSection_vs_E.dat};

\end{semilogyaxis}
\end{tikzpicture}
\caption{Production cross-section of the $\phi$ boson, with mass $m_{\phi} = m_{\tau}$ and coupling constant $|g_{V}|=3\times 10^{-3}$, as a function of the incoming muon energy. }  \label{fig:Crosssection}
\end{figure}

To calculate the lifetime of the $\phi$ boson, $\tau_{\phi}$, we use Equations 3.2-3.5 in \cite{Pospelov23} adapted to the $(\mu, \tau)$ case. We calculate the cross-section and decay-width using the GNU Scientific Library \cite{galassi2018scientific}. The production cross-section for the $\phi$ boson as a function of the incoming lepton energy is shown in Fig.~\ref{fig:Crosssection}. Assuming $m_{\phi} \simeq m_{\tau}$, the threshold for the production is given by $E_{\mu} > [(2m_{\tau} + m_{N})^{2} - m_{\mu}^{2} - m_{N}^{2}]/2m_{N} \simeq 3.8$ GeV for Pb, above which the cross-section steeply rises.
In the following, we use the corresponding muon beam flux of each experiment (NA64$\mu$ and DUNE) to evaluate the expressions for the $\phi$ production via the variables defined in Eq.~\ref{Eq:Mandelstam}.

\section{\label{sec:Simulation}$\mu$-on-target experiments}

We estimate the projected sensitivity of an experiment by finding the pair of parameter values $(g_{V}, m_\phi)$, where $N_\phi$ appropriate signal events are produced either directly in the target (NA64$\mu$) or in the detector (DUNE), as explained later, after a given exposure. In this work we exploit the $\phi$ boson production process using different and complementary $\mu$-on-target experiments: NA64$\mu$ with the active dump technique compared to proton beam-dump experiments such as the DUNE neutrino experiments using LBNF proton beam. Although the underlying production mechanism is the same, the two techniques differ in the flux of the muon beam, $\Phi_{\mu}(E)$, the target thickness and materials. 

In a general case, $\phi$ bosons are generated by the $\mu$-on-target process, and, after production, a fraction of them decay inside a detector volume and can be detected. The number of such signal events is
\begin{equation}\label{Eq:MuOnTarget1a}
N_{\phi} = \int dE_{\phi} \Phi_{\phi}(E_\phi) \times \frac{l_{\mathrm{det}}}{\gamma \beta c \tau_{\phi}},
\end{equation}
where $\gamma$ is the relativistic Lorentz-factor, and $\Phi_{\phi}(E_\phi)$ is the flux of the $\phi$ boson at the detector, estimated as
\begin{align}\label{Eq:MuOnTarget1b}
& \Phi_{\phi}(E_{\phi}) = \int dE\Phi_{\mu}(E) \times \\ \nonumber
& \int_{E_{\mathrm{min}}}^{E} dE_{l}\frac{n_A}{-dE/dl}\int_{0}^{\theta_{\mathrm{det}}}d\theta_{\phi}\sin\theta_{\phi}\frac{d^{2}\sigma(E_l, E_\phi)}{dE_{\phi}d\cos\theta_{\phi}}.
\end{align}

Here, $\Phi_{\mu}(E)$ is the flux of the muon beam as a function of energy, $n_{A}$ is the number of target atoms per volume, $E_{l}$ is the muon energy after traveling a length $l$ in the target and losing energy according to the stopping power $-dE/dl$, $E_\mathrm{min}$ is the energy of the muon at the end of the target, and $\theta_{\mathrm{det}}$ is the angular acceptance.

In the next subsections we separately describe the two experimental scenarios and the assumptions in the derived sensitivity limits.
\subsection{NA64$\mu$ scenario}

NA64$\mu$ is a fixed-target experiment at CERN looking for new particles of Dark Matter and portal interactions produced in electromagnetic showers and coupled to muons. The experiment uses the secondary 160 GeV muons from the interactions of 400 GeV protons from the CERN SPS with a target. A set of beam scintillators and veto counters, low material-budget trackers and dipole magnets allow to precisely constrain the momentum of the incoming 160-GeV muons impinging on an active target. The main detector, where $\phi$ may be produced, consists of an electromagnetic calorimeter with 40 $X_0$ radiation length. Downstream, the detector is further equipped with veto counters and a $\sim$30-interaction length hadronic calorimeter. New particles could be produced by the muon beam scattering in the target and decay later to visible SM particles that could be seen by their signatures in a downstream detector. The current work is based on the detection of missing energy and momentum carried away by the produced hypothetical, long-lived $\phi$ boson, leaving a scattered muon as experimental signature (the momentum of the scattered muon ranging between $10-80$ GeV/c). The sensitivity in the search for the $\phi$ boson is higher with respect to the beam-dump approach due to the lower power in the coupling strength without a decay vertex. Thus, in NA64$\mu$ only the number of events at the production target needs to be estimated. Therefore, the number of events is given by $N_{\phi} = \int dE_{\phi} \Phi_{\phi}(E_\phi)$. Furthermore, the production target thickness is small and the muon energy loss can be neglected. As a result we use the following expression to estimate the $\phi$ boson flux,
\begin{align}\label{Eq:MuOnTarget2}
  & \Phi_{\phi}(E_{\phi}) = l_\mathrm{target} n_{A}\int dE\Phi_{\mu}(E) \times \\ \nonumber
& \int_{0}^{\theta_{\mathrm{det}}}d\theta_{\phi}\sin\theta_{\phi}\frac{d^{2}\sigma(E, E_\phi)}{dE_{\phi}d\cos\theta_{\phi}},
\end{align}  
where $l_\mathrm{target}$ is the thickness of the target. 

There has been an extensive study of simulating and evaluating the production of Dark Matter particles with a muon beam at NA64 \cite{BONDI2021108129, PhysRevD.104.076012, sieber2023probing}. We use the same method in this work to estimate the sensitivity reach of the experiment. We assume a muon beam with a mean of 160 GeV energy and a width of 4.3 GeV hitting a lead target with a total data of $\sim3\times 10^{13}$ muons-on-target (MOT). 

\subsection{\label{sec:DUNE}DUNE/LBNF scenario}

\begin{figure}[h]
\begin{tikzpicture}[scale=0.9]
\begin{semilogyaxis}[
    xlabel={$E_{\mu}$ [GeV]},
    ylabel={$\Phi_{\mu}(E$) [a.u.]},
    xmin=0, xmax=80,
    ymin=1, ymax=100000,
  ytick={0.1, 1,10,10^2, 10^3, 10^4, 10^5},
    minor tick num = 1,
    width = 0.45\textwidth,   
    height = 0.5\textwidth,
    legend pos=north east,
    ymajorgrids=false,
    grid style=dashed,
]


\addplot+[black, mark options={fill=black}, only marks, error bars/.cd,
  y dir=both, y explicit] table [y error index=2]{Figs/MuFlux.dat};

\end{semilogyaxis}
\end{tikzpicture}
\caption{Energy spectrum of muons at the end of the decay pipe from the full DUNE neutrino beam line simulation \cite{abi2020deep}, see explanation in the text. }  \label{fig:DUNEmu}
\end{figure}
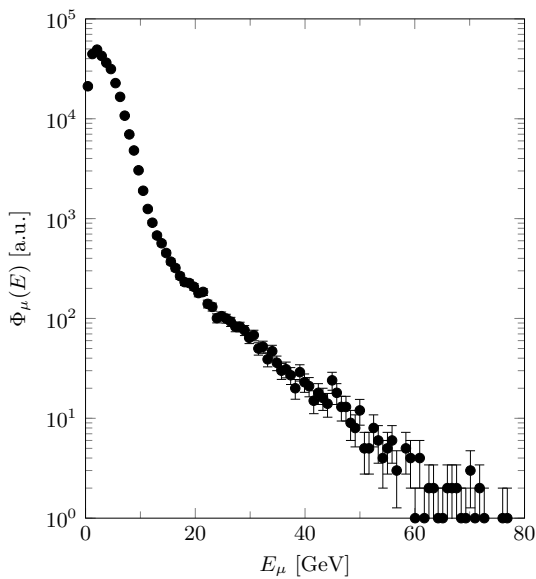

The Deep Underground Neutrino Experiment is a next-generation, wide-energy beam long-baseline neutrino experiment at Fermilab. It will use an intense (anti)neutrino beam that passes through a Near Detector at Fermilab and a Far Detector 1300 km away in South Dakota. The neutrino beam line of DUNE is a result of an exhaustive design and optimization work \cite{abi2020deep}. The beam is produced by a 60-120 GeV proton beam hitting a graphite target after which the produced pions and kaons decay to leptons and neutrinos in a $\sim 220 $-m-long decay pipe. At the end of pipe a dedicated $\sim 30$-m-long stainless-steel structure acts as a  beam dump to stop all muons $300$ m upstream from the Near Detector. We use the simulation of the neutrino beam production to trace particles along the beam axis. The muon flux used in the calculation is estimated from a dedicated tracking plane, which is located at the end of the decay pipe in the simulation. An example of the obtained muon energy spectrum is illustrated in Fig.~\ref{fig:DUNEmu}. The peak of the spectrum is at $E_{\mu} \simeq 2.5$ GeV, however, the long high-energy tail is responsible for the majority of $\phi$ bosons production due to the rapid increase of the cross-section with the incoming lepton energy (see Fig.~\ref{fig:Crosssection}). At the lower energy region the production rate is much smaller. From the neutrino flux simulation we estimate an integrated muon flux of $\Phi_{\mu} \simeq 5\times 10^{19}$ muons for $1.1\times 10^{21}$ Proton-On-Target (POT), corresponding to one year of data taking. 

In DUNE, a signal could be detected in the Near Detector from the decay of the $\phi$ boson that was produced by the muons hitting the stainless-steel dump . We consider the decay channel $\phi \rightarrow \mu^{+}\mu^{-}\nu_{\mu}\nu_{\tau}$ with a branching ratio of $17\%$ \cite{Pospelov23}, motivated by the dimuon results from NuTeV \cite{PhysRevLett.87.041801}. Possible backgrounds leading to a dimuon signature include deep inelastic scattering (DIS) and resonance production of mesons in charged-current (CC) muon-neutrino ($\nu_\mu$) interactions with a target nucleus. The mesons could decay in semi-leptonic mode, producing an extra muon. In order to analyze such potential background processes we performed simulations with the GENIE \cite{GENIE} Monte Carlo (MC) event generator that provides a comprehensive neutrino interaction modeling in the $E_{\nu} \sim 100$ MeV $-$ few 100 GeV neutrino energy region, including quasi-elastic, resonance and DIS processes. Unlike the $\phi$ boson decay, events with DIS or resonant meson production are accompanied with additional activity in the final state. Similarly to previous findings for NuTeV, after discriminating for low-multiplicity events with two muons, we found that these backgrounds could be completely suppressed in a MC simulation of 400 million events of neutrino interactions on an Argon target (corresponding to $\sim5$ years of operation of DUNE at the nominal intensity). However, further studies are planned with a full detector simulation to get a detailed understanding of the possible bounds on the background rejection.

\section{Results}

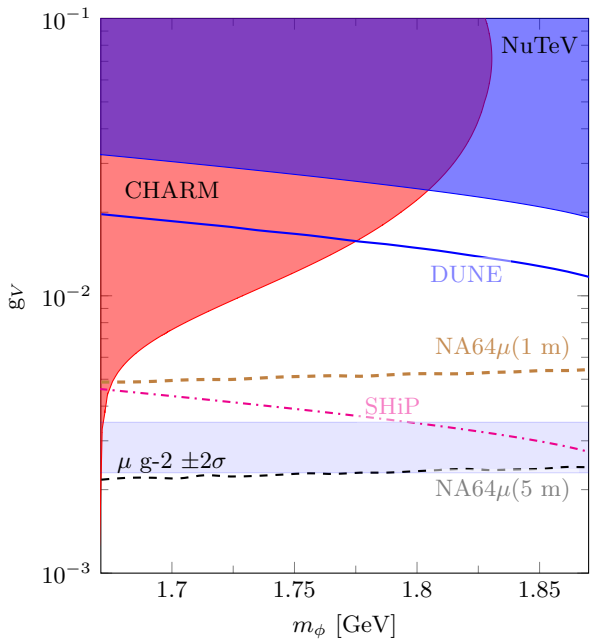
\begin{figure}
\begin{tikzpicture}
\begin{semilogyaxis}[
    xlabel={$m_{\phi}$ [GeV]},
    ylabel={g\(_V\)},
    xmin=1.671, xmax=1.87,
    ymin=1e-03, ymax=0.1,
    xtick={1.70,1.75,1.80,1.85},
    ytick={1e-04, 1e-03,1e-02,1e-01},
    minor tick num = 1,
    width = 0.45\textwidth,   
    height = 0.5\textwidth,
    legend pos=north east,
    ymajorgrids=false,
    grid style=dashed,
]

\addplot [
    smooth,
    thin,
    fill=red,
    fill opacity=0.5,
    color=red
    ] table {Figs/CHARM.dat};

\addplot [
    smooth,
    thin,
    fill=blue,
    fill opacity=0.5,
    color=blue
    ] table {Figs/NuTeV.dat};

\addplot [
    smooth,
    thick,
   dashdotted,
    color=magenta
    ] table {Figs/SHiP_muOT.dat} node [fill=white, opacity=0.5]
        at (axis cs:1.79,0.0040)   {SHiP};

\addplot[
    smooth,
    thick,
   solid,
    color=blue
    ]
    coordinates {
(1.6712, 0.0196846)
(1.68177, 0.0192718)
(1.69233, 0.0188677)
(1.7029, 0.018472)
(1.71347, 0.0180847)
(1.72403, 0.0177054)
(1.7346, 0.0172425)
(1.74516, 0.016881)
(1.75573, 0.016527)
(1.76629, 0.0160949)
(1.77686, 0.0156741)
(1.78743, 0.0153454)
(1.79799, 0.0149442)
(1.80856, 0.0145535)
(1.81912, 0.0140981)
(1.82969, 0.013657)
(1.84025, 0.0132297)
(1.85082, 0.012748)
(1.86139, 0.012219)
(1.87195, 0.0115884)
     } node [fill=white, opacity=0.5]
       at (axis cs:1.82,0.012)   {DUNE};



\addplot[
    smooth,
    very thick,
    dashed,
    color=brown
    ]coordinates {
    (1.6712, 0.00488603)
(1.68177, 0.00488603)
(1.69233, 0.00491199)
(1.7029, 0.00496431)
(1.71347, 0.00499068)
(1.72403, 0.00499068)
(1.7346, 0.00504385)
(1.74516, 0.00509758)
(1.75573, 0.00512466)
(1.76629, 0.00515189)
(1.77686, 0.00512466)
(1.78743, 0.00517926)
(1.79799, 0.00523443)
(1.80856, 0.00523443)
(1.81912, 0.00523443)
(1.82969, 0.00529019)
(1.84025, 0.0053183)
(1.85082, 0.00537495)
(1.86139, 0.00537495)
(1.87195, 0.00543221)
    } node [fill=white, opacity=0.9]
        at (axis cs:1.835,0.0065)   {NA64$\mu$(1 m)};

\addplot[
    smooth,
    thick,
    dashed,
    color=black
    ]coordinates {
(1.6712, 0.00217219)
(1.68177, 0.002207)
(1.69233, 0.002207)
(1.7029, 0.00219533)
(1.71347, 0.00225427)
(1.72403, 0.00223051)
(1.7346, 0.00224236)
(1.74516, 0.00226624)
(1.75573, 0.00229039)
(1.76629, 0.00227828)
(1.77686, 0.00230255)
(1.78743, 0.00230255)
(1.79799, 0.00232708)
(1.80856, 0.00235187)
(1.81912, 0.00237693)
(1.82969, 0.00235187)
(1.84025, 0.00236437)
(1.85082, 0.00237693)
(1.86139, 0.00241501)
(1.87195, 0.00240225)
    } node [fill=white, opacity=0.5]
        at (axis cs:1.835,0.002)   {NA64$\mu$(5 m)};

\addplot[name path=f,domain=1.66:1.89,blue,opacity=0.15] {3.5e-03};
\addplot[name path=f2,domain=1.66:1.89,blue,opacity=0.15] {2.3e-03};
\addplot [
        thick,
        color=blue,
        fill=blue, 
        fill opacity=0.1
    ]
    fill between[
        of=f and f2,
        soft clip={domain=1.66:1.89},
    ];
    \node [rotate=0] at (axis cs:  1.7,  2.5e-03) {$\mu$ g-2 $\pm 2\sigma$ };

\node [rotate=0] at (axis cs:  1.7,  0.024) {CHARM};

\node [rotate=0] at (axis cs:  1.85,  0.08) {NuTeV};

    
\end{semilogyaxis}
\end{tikzpicture}
\caption{Constraints from CHARM, NuTeV  \cite{PhysRevLett.87.041801, DORENBOSCH1986473}, and projected sensitivity curves for SHiP \cite{Pospelov23, Alekhin_2016} (dash-dotted, magenta line), DUNE (solid, blue line) and NA64$\mu$ (dashed, brown and black lines), in the $\mu$-on-target production mode (except for CHARM). In the case of NA64$\mu$ the optimized setup with 1-m-long (brown dashed line) and 5-m-long (black dashed line) targets are also shown separately.
The light blue region shows the benchmark parameter range explaining the muon $g_{\mu}-2$ anomaly. }  
\label{fig:Limits}
\end{figure}

We illustrate the sensitivity potential for the benchmark CLFV scenario with the complementary muon beams at NA64$\mu$ and DUNE in Fig~\ref{fig:Limits}. For both experiments, the double-differential cross-section in Eq.~\ref{Eq:MuOnTarget1b} or \ref{Eq:MuOnTarget2} is evaluated given the energy spectra of each experiment, $\Phi_{\mu}(E)$, and the kinematical limits on the final-state $\phi$-boson fractional energy, $x$, which is constrainted by the masses of the boson $\phi$ and the $\tau$.

In the case of the NA64$\mu$ experiment a total integrated muon flux of $\sim3\times10^{13}$ MOT is achievable \cite{Gninenko:2653581}. The time needed to accumulate the assumed total MOT is estimated to be $\sim 100$ days. This conservative estimation is based on the CERN SPS delivering on average 3500 spills per day and $2\times 10^{8}$ muons per spill. Benefiting from the unique combination of a 160-GeV muon beam with missing-energy and momentum search, NA64$\mu$ would be able to perform a competitive search for such a CLFV signal. Furthermore, the sensitivity reach critically depends on the length of the active target. We find that a 1-m-long target would already be able to explore a large part of the parameter space, $g_{V} \geq 6\times 10^{-3}$. However, the experiment is highly modular and a possible optimization of the setup could enhance its potential further: a feasible option is to increase the target length to 5 meters which would allow to completely cover the $(g_{\mu}-2)$ preferred region and to probe a variety of other new physics scenarios involving muons. We also find that already with a $10^{12}$ MOT and a $\sim 3$ m long target the $g_{V} \geq \times 10^{-2}$ parameter region can be covered. A detailed Monte Carlo simulation of an optimized setup will follow as a next step.

For the DUNE experiment the $\phi$-boson production is driven by the high-end tail of the muon flux. Compared to NA64$\mu$, the lower muon energies at DUNE and thus lower production cross-section are compensated by the more intense muon flux. Assuming 20 years of operation at nominal intensity, DUNE would be able to explore a significant part of the parameter space, reaching into the $g_{V} \simeq 10^{-2}$ region and potentially improving the constraints from NuTeV. However, an optimization of the neutrino beam line could further enhance the contribution of high-energy muons in the flux and subsequently approach the $g_{V} \leq 10^{-2}$ benchmark region. This scenario is partially motivated by a recent work exploring an alternative beam-dump operation mode to probe new physics with DUNE \cite{PhysRevD.107.055043}. 

For comparison we also show the projected sensitivity for the same $\mu$-on-target mode calculated by \cite{Pospelov23} for CHARM, NuTeV and SHiP. Here we do not show the constraints or projected sensitivity limits derived for the direct electroweak and heavy meson decay process, which was included in the previous work \cite{Pospelov23} since we only focus on the $\mu$-on-target scenario. In the case of SHiP, the assumed total data corresponds to $2\times 10^{20}$ protons-on-target (POT). The worse sensitivity in DUNE compared to SHiP stems from the lower muon energies in the flux \cite{Ahdida_2020}. It is also noted that there are differences in the beam intensity between NA64$\mu$ and SHiP. We assume a $10^{12}$ POT per spill intensity at the CERN SPS in the case of the muon beam line used by NA64$\mu$, while for SHiP the proton beam intensity is usually expected to be an order of magnitude higher. 

We note that a similar setup of NA64$\mu$ is capable of searching for other new scalar particle candidates using the same muon beam \cite{sieber2023probing}. In addition, the proposed Muon Missing Momentum ($M^{3}$) experiment at Fermilab \cite{Kahn:2018cqs} also plans to probe new physics with a dedicated muon beam. Finally, a number of experiments also have the potential to search for hidden-sector scalar particles, such as  SHADOWS \cite{ShadowsLOI}, HIKE \cite{HIKE:2022qra}, and ATLAS \cite{Galon:2019owl}.


\section{Conclusions}

In summary, we present the sensitivity potential of two $\mu$-on-target experiments: NA64$\mu$ and DUNE, as complementary modes of searching for new physics with muon beams. We find that both NA64$\mu$ and DUNE have the potential to cover a significant portion of the benchmark model parameter space, ($m_{\phi}, g_{V}$). NA64$\mu$ with an optimized setup could probe the coupling parameter down to $g_{V}\simeq 3\times 10^{-3}$, completely covering the muon $g_{\mu} - 2$ preferred region and thus providing a similar projected reach as SHiP. DUNE will also be able to cover unexplored parts of the parameter space, potentially improving on the obtained constraints from NuTeV. An optimization of the neutrino beam line, increasing the contribution from the high-energy tail, could allow to further enhance the sensitivity of DUNE.

Although we use a given CLFV model as a benchmark in this work, we note that similar techniques can be used to study the sensitivity potential of experiments with muon beams to other physics scenarios. Beyond the $g_{\mu}-2$ discrepancy and neutrino flavor oscillations, searches for Dark Sector particles are also motivated by the matter-antimatter asymmetry, the known Dark Matter abundance from astrophysical and cosmological observations, or by theoretical motivations strongly suggesting the existence of additional gauge groups weakly coupling to SM fields \cite{doi:10.1146/annurev-nucl-102419-055056, 2023arXiv230501715A}.

\begin{acknowledgments}
We gratefully acknowledge conversations with D. Harris. The work of LMB is supported by SNSF Grant No. 186158 (Switzerland), RyC-030551-I, and PID2021-123955NA-100 funded by MCIN/AEI/ 10.13039/501100011033/FEDER, UE (Spain).
\end{acknowledgments}


\bibliography{Bibliography}

\begin{thebibliography}{40}%
\makeatletter
\providecommand \@ifxundefined [1]{%
 \@ifx{#1\undefined}
}%
\providecommand \@ifnum [1]{%
 \ifnum #1\expandafter \@firstoftwo
 \else \expandafter \@secondoftwo
 \fi
}%
\providecommand \@ifx [1]{%
 \ifx #1\expandafter \@firstoftwo
 \else \expandafter \@secondoftwo
 \fi
}%
\providecommand \natexlab [1]{#1}%
\providecommand \enquote  [1]{``#1''}%
\providecommand \bibnamefont  [1]{#1}%
\providecommand \bibfnamefont [1]{#1}%
\providecommand \citenamefont [1]{#1}%
\providecommand \href@noop [0]{\@secondoftwo}%
\providecommand \href [0]{\begingroup \@sanitize@url \@href}%
\providecommand \@href[1]{\@@startlink{#1}\@@href}%
\providecommand \@@href[1]{\endgroup#1\@@endlink}%
\providecommand \@sanitize@url [0]{\catcode `\\12\catcode `\$12\catcode
  `\&12\catcode `\#12\catcode `\^12\catcode `\_12\catcode `\%12\relax}%
\providecommand \@@startlink[1]{}%
\providecommand \@@endlink[0]{}%
\providecommand \url  [0]{\begingroup\@sanitize@url \@url }%
\providecommand \@url [1]{\endgroup\@href {#1}{\urlprefix }}%
\providecommand \urlprefix  [0]{URL }%
\providecommand \Eprint [0]{\href }%
\providecommand \doibase [0]{http://dx.doi.org/}%
\providecommand \selectlanguage [0]{\@gobble}%
\providecommand \bibinfo  [0]{\@secondoftwo}%
\providecommand \bibfield  [0]{\@secondoftwo}%
\providecommand \translation [1]{[#1]}%
\providecommand \BibitemOpen [0]{}%
\providecommand \bibitemStop [0]{}%
\providecommand \bibitemNoStop [0]{.\EOS\space}%
\providecommand \EOS [0]{\spacefactor3000\relax}%
\providecommand \BibitemShut  [1]{\csname bibitem#1\endcsname}%
\let\auto@bib@innerbib\@empty
\bibitem [{\citenamefont {Lanfranchi}\ \emph {et~al.}(2021)\citenamefont
  {Lanfranchi}, \citenamefont {Pospelov},\ and\ \citenamefont
  {Schuster}}]{doi:10.1146/annurev-nucl-102419-055056}%
  \BibitemOpen
  \bibfield  {author} {\bibinfo {author} {\bibfnamefont {G.}~\bibnamefont
  {Lanfranchi}}, \bibinfo {author} {\bibfnamefont {M.}~\bibnamefont
  {Pospelov}}, \ and\ \bibinfo {author} {\bibfnamefont {P.}~\bibnamefont
  {Schuster}},\ }\href {\doibase 10.1146/annurev-nucl-102419-055056} {\bibfield
   {journal} {\bibinfo  {journal} {Annual Review of Nuclear and Particle
  Science}\ }\textbf {\bibinfo {volume} {71}},\ \bibinfo {pages} {279}
  (\bibinfo {year} {2021})}\BibitemShut {NoStop}%
\bibitem [{\citenamefont {Abi}\ \emph {et~al.}(2021)\citenamefont {Abi} \emph
  {et~al.}}]{PhysRevLett.126.141801}%
  \BibitemOpen
  \bibfield  {author} {\bibinfo {author} {\bibfnamefont {B.}~\bibnamefont
  {Abi}} \emph {et~al.} (\bibinfo {collaboration} {Muon $g\ensuremath{-}2$
  Collaboration}),\ }\href {\doibase 10.1103/PhysRevLett.126.141801} {\bibfield
   {journal} {\bibinfo  {journal} {Phys. Rev. Lett.}\ }\textbf {\bibinfo
  {volume} {126}},\ \bibinfo {pages} {141801} (\bibinfo {year}
  {2021})}\BibitemShut {NoStop}%
\bibitem [{\citenamefont {Aoyama}\ \emph {et~al.}(2020)\citenamefont {Aoyama}
  \emph {et~al.}}]{AOYAMA20201}%
  \BibitemOpen
  \bibfield  {author} {\bibinfo {author} {\bibfnamefont {T.}~\bibnamefont
  {Aoyama}} \emph {et~al.},\ }\href {\doibase
  https://doi.org/10.1016/j.physrep.2020.07.006} {\bibfield  {journal}
  {\bibinfo  {journal} {Physics Reports}\ }\textbf {\bibinfo {volume} {887}},\
  \bibinfo {pages} {1} (\bibinfo {year} {2020})}\BibitemShut {NoStop}%
\bibitem [{\citenamefont {Borsanyi}\ \emph {et~al.}(2021)\citenamefont
  {Borsanyi} \emph {et~al.}}]{BorsanyiNATURE}%
  \BibitemOpen
  \bibfield  {author} {\bibinfo {author} {\bibfnamefont {S.}~\bibnamefont
  {Borsanyi}} \emph {et~al.},\ }\href@noop {} {\bibfield  {journal} {\bibinfo
  {journal} {Nature}\ }\textbf {\bibinfo {volume} {593}},\ \bibinfo {pages}
  {51} (\bibinfo {year} {2021})}\BibitemShut {NoStop}%
\bibitem [{\citenamefont {Chen}\ \emph {et~al.}(2017)\citenamefont {Chen},
  \citenamefont {Pospelov},\ and\ \citenamefont {Zhong}}]{PhysRevD.95.115005}%
  \BibitemOpen
  \bibfield  {author} {\bibinfo {author} {\bibfnamefont {C.-Y.}\ \bibnamefont
  {Chen}}, \bibinfo {author} {\bibfnamefont {M.}~\bibnamefont {Pospelov}}, \
  and\ \bibinfo {author} {\bibfnamefont {Y.-M.}\ \bibnamefont {Zhong}},\ }\href
  {\doibase 10.1103/PhysRevD.95.115005} {\bibfield  {journal} {\bibinfo
  {journal} {Phys. Rev. D}\ }\textbf {\bibinfo {volume} {95}},\ \bibinfo
  {pages} {115005} (\bibinfo {year} {2017})}\BibitemShut {NoStop}%
\bibitem [{\citenamefont {Meucci}(2022)}]{Meucci:20229n}%
  \BibitemOpen
  \bibfield  {author} {\bibinfo {author} {\bibfnamefont {M.}~\bibnamefont
  {Meucci}},\ }\href {\doibase 10.22323/1.402.0120} {\bibfield  {journal}
  {\bibinfo  {journal} {PoS}\ }\textbf {\bibinfo {volume} {NuFact2021}},\
  \bibinfo {pages} {120} (\bibinfo {year} {2022})}\BibitemShut {NoStop}%
\bibitem [{\citenamefont {Chiappini}\ \emph {et~al.}(2021)\citenamefont
  {Chiappini} \emph {et~al.}}]{universe7120466}%
  \BibitemOpen
  \bibfield  {author} {\bibinfo {author} {\bibfnamefont {M.}~\bibnamefont
  {Chiappini}} \emph {et~al.} (\bibinfo {collaboration} {{MEG II
  Collaboration}}),\ }\href {https://www.mdpi.com/2218-1997/7/12/466}
  {\bibfield  {journal} {\bibinfo  {journal} {Universe}\ }\textbf {\bibinfo
  {volume} {7}} (\bibinfo {year} {2021})}\BibitemShut {NoStop}%
\bibitem [{\citenamefont {Bartoszek}\ \emph {et~al.}(2014)\citenamefont
  {Bartoszek} \emph {et~al.}}]{Mu2e:2014fns}%
  \BibitemOpen
  \bibfield  {author} {\bibinfo {author} {\bibfnamefont {L.}~\bibnamefont
  {Bartoszek}} \emph {et~al.} (\bibinfo {collaboration} {Mu2e}),\ }\href
  {\doibase 10.2172/1172555} {\  (\bibinfo {year} {2014}),\ 10.2172/1172555},\
  \Eprint {http://arxiv.org/abs/1501.05241} {arXiv:1501.05241
  [physics.ins-det]} \BibitemShut {NoStop}%
\bibitem [{\citenamefont {Hesketh}\ \emph {et~al.}(2022)\citenamefont
  {Hesketh}, \citenamefont {Hughes}, \citenamefont {Perrevoort},\ and\
  \citenamefont {Rompotis}}]{Hesketh:2022wgw}%
  \BibitemOpen
  \bibfield  {author} {\bibinfo {author} {\bibfnamefont {G.}~\bibnamefont
  {Hesketh}}, \bibinfo {author} {\bibfnamefont {S.}~\bibnamefont {Hughes}},
  \bibinfo {author} {\bibfnamefont {A.-K.}\ \bibnamefont {Perrevoort}}, \ and\
  \bibinfo {author} {\bibfnamefont {N.}~\bibnamefont {Rompotis}} (\bibinfo
  {collaboration} {Mu3e}),\ }in\ \href@noop {} {\emph {\bibinfo {booktitle}
  {{Snowmass 2021}}}}\ (\bibinfo {year} {2022})\ \Eprint
  {http://arxiv.org/abs/2204.00001} {arXiv:2204.00001 [hep-ex]} \BibitemShut
  {NoStop}%
\bibitem [{\citenamefont {Abramishvili}\ \emph {et~al.}(2020)\citenamefont
  {Abramishvili} \emph {et~al.}}]{COMET:2018auw}%
  \BibitemOpen
  \bibfield  {author} {\bibinfo {author} {\bibfnamefont {R.}~\bibnamefont
  {Abramishvili}} \emph {et~al.} (\bibinfo {collaboration} {COMET}),\ }\href
  {\doibase 10.1093/ptep/ptz125} {\bibfield  {journal} {\bibinfo  {journal}
  {PTEP}\ }\textbf {\bibinfo {volume} {2020}},\ \bibinfo {pages} {033C01}
  (\bibinfo {year} {2020})},\ \Eprint {http://arxiv.org/abs/1812.09018}
  {arXiv:1812.09018 [physics.ins-det]} \BibitemShut {NoStop}%
\bibitem [{\citenamefont {Altmannshofer}\ \emph {et~al.}(2019)\citenamefont
  {Altmannshofer} \emph {et~al.}}]{Belle-II:2018jsg}%
  \BibitemOpen
  \bibfield  {author} {\bibinfo {author} {\bibfnamefont {W.}~\bibnamefont
  {Altmannshofer}} \emph {et~al.} (\bibinfo {collaboration} {Belle-II}),\
  }\href {\doibase 10.1093/ptep/ptz106} {\bibfield  {journal} {\bibinfo
  {journal} {PTEP}\ }\textbf {\bibinfo {volume} {2019}},\ \bibinfo {pages}
  {123C01} (\bibinfo {year} {2019})},\ \bibinfo {note} {[Erratum: PTEP 2020,
  029201 (2020)]},\ \Eprint {http://arxiv.org/abs/1808.10567} {arXiv:1808.10567
  [hep-ex]} \BibitemShut {NoStop}%
\bibitem [{\citenamefont {Abud}\ \emph {et~al.}(2022)\citenamefont {Abud} \emph
  {et~al.}}]{PhysRevD.105.072006}%
  \BibitemOpen
  \bibfield  {author} {\bibinfo {author} {\bibfnamefont {A.~A.}\ \bibnamefont
  {Abud}} \emph {et~al.} (\bibinfo {collaboration} {{DUNE Collaboration}}),\
  }\href {\doibase 10.1103/PhysRevD.105.072006} {\bibfield  {journal} {\bibinfo
   {journal} {Phys. Rev. D}\ }\textbf {\bibinfo {volume} {105}},\ \bibinfo
  {pages} {072006} (\bibinfo {year} {2022})}\BibitemShut {NoStop}%
\bibitem [{\citenamefont {{Abe}}\ \emph {et~al.}(2015)\citenamefont {{Abe}}
  \emph {et~al.}}]{2015PTEP.2015e3C02A}%
  \BibitemOpen
  \bibfield  {author} {\bibinfo {author} {\bibfnamefont {K.}~\bibnamefont
  {{Abe}}} \emph {et~al.},\ }\href {\doibase 10.1093/ptep/ptv061} {\bibfield
  {journal} {\bibinfo  {journal} {Progress of Theoretical and Experimental
  Physics}\ }\textbf {\bibinfo {volume} {2015}},\ \bibinfo {eid} {053C02}
  (\bibinfo {year} {2015})}\BibitemShut {NoStop}%
\bibitem [{\citenamefont {{Y. Ema, Z. Liu, K-F Lyu and M.
  Pospelov}}(2023)}]{Pospelov23}%
  \BibitemOpen
  \bibfield  {author} {\bibinfo {author} {\bibnamefont {{Y. Ema, Z. Liu, K-F
  Lyu and M. Pospelov}}},\ }\href@noop {} {\bibfield  {journal} {\bibinfo
  {journal} {{JHEP}}\ }\textbf {\bibinfo {volume} {135}} (\bibinfo {year}
  {2023})}\BibitemShut {NoStop}%
\bibitem [{\citenamefont {Auerbach}\ \emph {et~al.}(2001)\citenamefont
  {Auerbach} \emph {et~al.}}]{PhysRevD.63.112001}%
  \BibitemOpen
  \bibfield  {author} {\bibinfo {author} {\bibfnamefont {L.~B.}\ \bibnamefont
  {Auerbach}} \emph {et~al.} (\bibinfo {collaboration} {LSND Collaboration}),\
  }\href {\doibase 10.1103/PhysRevD.63.112001} {\bibfield  {journal} {\bibinfo
  {journal} {Phys. Rev. D}\ }\textbf {\bibinfo {volume} {63}},\ \bibinfo
  {pages} {112001} (\bibinfo {year} {2001})}\BibitemShut {NoStop}%
\bibitem [{\citenamefont {Adams}\ \emph {et~al.}(2001)\citenamefont {Adams}
  \emph {et~al.}}]{PhysRevLett.87.041801}%
  \BibitemOpen
  \bibfield  {author} {\bibinfo {author} {\bibfnamefont {T.}~\bibnamefont
  {Adams}} \emph {et~al.},\ }\href {\doibase 10.1103/PhysRevLett.87.041801}
  {\bibfield  {journal} {\bibinfo  {journal} {Phys. Rev. Lett.}\ }\textbf
  {\bibinfo {volume} {87}},\ \bibinfo {pages} {041801} (\bibinfo {year}
  {2001})}\BibitemShut {NoStop}%
\bibitem [{\citenamefont {Dorenbosch}\ \emph {et~al.}(1986)\citenamefont
  {Dorenbosch} \emph {et~al.}}]{DORENBOSCH1986473}%
  \BibitemOpen
  \bibfield  {author} {\bibinfo {author} {\bibfnamefont {J.}~\bibnamefont
  {Dorenbosch}} \emph {et~al.},\ }\href {\doibase
  https://doi.org/10.1016/0370-2693(86)91601-1} {\bibfield  {journal} {\bibinfo
   {journal} {Physics Letters B}\ }\textbf {\bibinfo {volume} {166}},\ \bibinfo
  {pages} {473} (\bibinfo {year} {1986})}\BibitemShut {NoStop}%
\bibitem [{Ale(2016)}]{Alekhin_2016}%
  \BibitemOpen
  \href {\doibase 10.1088/0034-4885/79/12/124201} {\bibfield  {journal}
  {\bibinfo  {journal} {Reports on Progress in Physics}\ }\textbf {\bibinfo
  {volume} {79}},\ \bibinfo {pages} {124201} (\bibinfo {year}
  {2016})}\BibitemShut {NoStop}%
\bibitem [{\citenamefont {Abi}\ \emph {et~al.}(2020)\citenamefont {Abi} \emph
  {et~al.}}]{abi2020deep}%
  \BibitemOpen
  \bibfield  {author} {\bibinfo {author} {\bibfnamefont {B.}~\bibnamefont
  {Abi}} \emph {et~al.},\ }\href@noop {} {\enquote {\bibinfo {title} {{Deep
  Underground Neutrino Experiment (DUNE), Far Detector Technical Design Report,
  Volume II: DUNE Physics}},}\ } (\bibinfo {year} {2020}),\ \Eprint
  {http://arxiv.org/abs/2002.03005} {arXiv:2002.03005 [hep-ex]} \BibitemShut
  {NoStop}%
\bibitem [{\citenamefont {Gninenko}(2019)}]{Gninenko:2653581}%
  \BibitemOpen
  \bibfield  {author} {\bibinfo {author} {\bibfnamefont {S.}~\bibnamefont
  {Gninenko}} (\bibinfo {collaboration} {NA64}),\ }\href
  {https://cds.cern.ch/record/2653581} {\emph {\bibinfo {title} {{ Proposal for
  an experiment to search for dark sector particles weakly coupled to muon at
  the SPS}}}},\ \bibinfo {type} {Tech. Rep.}\ (\bibinfo  {institution} {CERN},\
  \bibinfo {address} {Geneva},\ \bibinfo {year} {2019})\BibitemShut {NoStop}%
\bibitem [{\citenamefont {Gninenko}\ \emph
  {et~al.}(2019{\natexlab{a}})\citenamefont {Gninenko} \emph
  {et~al.}}]{GninenkoNA64mu}%
  \BibitemOpen
  \bibfield  {author} {\bibinfo {author} {\bibfnamefont {S.~N.}\ \bibnamefont
  {Gninenko}} \emph {et~al.},\ }\href
  {https://doi.org/10.1016/j.physletb.2019.07.015} {\bibfield  {journal}
  {\bibinfo  {journal} {Physics Letters B}\ }\textbf {\bibinfo {volume}
  {796}},\ \bibinfo {pages} {117} (\bibinfo {year}
  {2019}{\natexlab{a}})}\BibitemShut {NoStop}%
\bibitem [{\citenamefont {Gninenko}\ \emph
  {et~al.}(2019{\natexlab{b}})\citenamefont {Gninenko}, \citenamefont
  {Kirpichnikov},\ and\ \citenamefont {Krasnikov}}]{Gninenko:2018ter}%
  \BibitemOpen
  \bibfield  {author} {\bibinfo {author} {\bibfnamefont {S.~N.}\ \bibnamefont
  {Gninenko}}, \bibinfo {author} {\bibfnamefont {D.~V.}\ \bibnamefont
  {Kirpichnikov}}, \ and\ \bibinfo {author} {\bibfnamefont {N.~V.}\
  \bibnamefont {Krasnikov}},\ }\href {\doibase 10.1103/PhysRevD.100.035003}
  {\bibfield  {journal} {\bibinfo  {journal} {Phys. Rev. D}\ }\textbf {\bibinfo
  {volume} {100}},\ \bibinfo {pages} {035003} (\bibinfo {year}
  {2019}{\natexlab{b}})},\ \Eprint {http://arxiv.org/abs/1810.06856}
  {arXiv:1810.06856 [hep-ph]} \BibitemShut {NoStop}%
\bibitem [{\citenamefont {Gninenko}\ \emph {et~al.}(2018)\citenamefont
  {Gninenko}, \citenamefont {Kovalenko}, \citenamefont {Kuleshov},
  \citenamefont {Lyubovitskij},\ and\ \citenamefont
  {Zhevlakov}}]{Gninenko:2018num}%
  \BibitemOpen
  \bibfield  {author} {\bibinfo {author} {\bibfnamefont {S.}~\bibnamefont
  {Gninenko}}, \bibinfo {author} {\bibfnamefont {S.}~\bibnamefont {Kovalenko}},
  \bibinfo {author} {\bibfnamefont {S.}~\bibnamefont {Kuleshov}}, \bibinfo
  {author} {\bibfnamefont {V.~E.}\ \bibnamefont {Lyubovitskij}}, \ and\
  \bibinfo {author} {\bibfnamefont {A.~S.}\ \bibnamefont {Zhevlakov}},\ }\href
  {\doibase 10.1103/PhysRevD.98.015007} {\bibfield  {journal} {\bibinfo
  {journal} {Phys. Rev. D}\ }\textbf {\bibinfo {volume} {98}},\ \bibinfo
  {pages} {015007} (\bibinfo {year} {2018})},\ \Eprint
  {http://arxiv.org/abs/1804.05550} {arXiv:1804.05550 [hep-ph]} \BibitemShut
  {NoStop}%
\bibitem [{\citenamefont {Gninenko}\ \emph
  {et~al.}(2019{\natexlab{c}})\citenamefont {Gninenko}, \citenamefont
  {Kirpichnikov}, \citenamefont {Kirsanov},\ and\ \citenamefont
  {Krasnikov}}]{Gninenko:2019qiv}%
  \BibitemOpen
  \bibfield  {author} {\bibinfo {author} {\bibfnamefont {S.~N.}\ \bibnamefont
  {Gninenko}}, \bibinfo {author} {\bibfnamefont {D.~V.}\ \bibnamefont
  {Kirpichnikov}}, \bibinfo {author} {\bibfnamefont {M.~M.}\ \bibnamefont
  {Kirsanov}}, \ and\ \bibinfo {author} {\bibfnamefont {N.~V.}\ \bibnamefont
  {Krasnikov}},\ }\href {\doibase 10.1016/j.physletb.2019.07.015} {\bibfield
  {journal} {\bibinfo  {journal} {Phys. Lett. B}\ }\textbf {\bibinfo {volume}
  {796}},\ \bibinfo {pages} {117} (\bibinfo {year} {2019}{\natexlab{c}})},\
  \Eprint {http://arxiv.org/abs/1903.07899} {arXiv:1903.07899 [hep-ph]}
  \BibitemShut {NoStop}%
\bibitem [{\citenamefont {Gninenko}\ and\ \citenamefont
  {Krasnikov}(2022)}]{Gninenko:2022ttd}%
  \BibitemOpen
  \bibfield  {author} {\bibinfo {author} {\bibfnamefont {S.~N.}\ \bibnamefont
  {Gninenko}}\ and\ \bibinfo {author} {\bibfnamefont {N.~V.}\ \bibnamefont
  {Krasnikov}} (\bibinfo {collaboration} {NA64}),\ }\href {\doibase
  10.1103/PhysRevD.106.015003} {\bibfield  {journal} {\bibinfo  {journal}
  {Phys. Rev. D}\ }\textbf {\bibinfo {volume} {106}},\ \bibinfo {pages}
  {015003} (\bibinfo {year} {2022})},\ \Eprint
  {http://arxiv.org/abs/2202.04410} {arXiv:2202.04410 [hep-ph]} \BibitemShut
  {NoStop}%
\bibitem [{\citenamefont {Sieber}\ \emph {et~al.}(2023)\citenamefont {Sieber},
  \citenamefont {Kirpichnikov}, \citenamefont {Voronchikhin}, \citenamefont
  {Crivelli}, \citenamefont {Gninenko}, \citenamefont {Kirsanov}, \citenamefont
  {Krasnikov}, \citenamefont {Molina-Bueno},\ and\ \citenamefont
  {Sekatskii}}]{sieber2023probing}%
  \BibitemOpen
  \bibfield  {author} {\bibinfo {author} {\bibfnamefont {H.}~\bibnamefont
  {Sieber}}, \bibinfo {author} {\bibfnamefont {D.~V.}\ \bibnamefont
  {Kirpichnikov}}, \bibinfo {author} {\bibfnamefont {I.~V.}\ \bibnamefont
  {Voronchikhin}}, \bibinfo {author} {\bibfnamefont {P.}~\bibnamefont
  {Crivelli}}, \bibinfo {author} {\bibfnamefont {S.~N.}\ \bibnamefont
  {Gninenko}}, \bibinfo {author} {\bibfnamefont {M.~M.}\ \bibnamefont
  {Kirsanov}}, \bibinfo {author} {\bibfnamefont {N.~V.}\ \bibnamefont
  {Krasnikov}}, \bibinfo {author} {\bibfnamefont {L.}~\bibnamefont
  {Molina-Bueno}}, \ and\ \bibinfo {author} {\bibfnamefont {S.~K.}\
  \bibnamefont {Sekatskii}},\ }\href@noop {} {\enquote {\bibinfo {title}
  {Probing hidden sectors with a muon beam: implication of spin-0 dark matter
  mediators for muon $(g-2)$ anomaly and validity of the weisz\"acker-williams
  approach},}\ } (\bibinfo {year} {2023}),\ \Eprint
  {http://arxiv.org/abs/2305.09015} {arXiv:2305.09015 [hep-ph]} \BibitemShut
  {NoStop}%
\bibitem [{\citenamefont {Bjorken}\ \emph {et~al.}(2009)\citenamefont
  {Bjorken}, \citenamefont {Essig}, \citenamefont {Schuster},\ and\
  \citenamefont {Toro}}]{PhysRevD.80.075018}%
  \BibitemOpen
  \bibfield  {author} {\bibinfo {author} {\bibfnamefont {J.~D.}\ \bibnamefont
  {Bjorken}}, \bibinfo {author} {\bibfnamefont {R.}~\bibnamefont {Essig}},
  \bibinfo {author} {\bibfnamefont {P.}~\bibnamefont {Schuster}}, \ and\
  \bibinfo {author} {\bibfnamefont {N.}~\bibnamefont {Toro}},\ }\href {\doibase
  10.1103/PhysRevD.80.075018} {\bibfield  {journal} {\bibinfo  {journal} {Phys.
  Rev. D}\ }\textbf {\bibinfo {volume} {80}},\ \bibinfo {pages} {075018}
  (\bibinfo {year} {2009})}\BibitemShut {NoStop}%
\bibitem [{\citenamefont {Inc.}()}]{Mathematica}%
  \BibitemOpen
  \bibfield  {author} {\bibinfo {author} {\bibfnamefont {W.~R.}\ \bibnamefont
  {Inc.}},\ }\href {https://www.wolfram.com/mathematica} {\enquote {\bibinfo
  {title} {Mathematica, {V}ersion 13.1},}\ }\bibinfo {note} {Champaign, IL,
  2022}\BibitemShut {NoStop}%
\bibitem [{\citenamefont {Shtabovenko}\ \emph {et~al.}(2020)\citenamefont
  {Shtabovenko}, \citenamefont {Mertig},\ and\ \citenamefont
  {Orellana}}]{SHTABOVENKO2020107478}%
  \BibitemOpen
  \bibfield  {author} {\bibinfo {author} {\bibfnamefont {V.}~\bibnamefont
  {Shtabovenko}}, \bibinfo {author} {\bibfnamefont {R.}~\bibnamefont {Mertig}},
  \ and\ \bibinfo {author} {\bibfnamefont {F.}~\bibnamefont {Orellana}},\
  }\href {\doibase https://doi.org/10.1016/j.cpc.2020.107478} {\bibfield
  {journal} {\bibinfo  {journal} {Computer Physics Communications}\ }\textbf
  {\bibinfo {volume} {256}},\ \bibinfo {pages} {107478} (\bibinfo {year}
  {2020})}\BibitemShut {NoStop}%
\bibitem [{\citenamefont {Galassi}\ \emph {et~al.}(2018)\citenamefont {Galassi}
  \emph {et~al.}}]{galassi2018scientific}%
  \BibitemOpen
  \bibfield  {author} {\bibinfo {author} {\bibfnamefont {M.}~\bibnamefont
  {Galassi}} \emph {et~al.},\ }\href {https://www.gnu.org/software/gsl/}
  {\enquote {\bibinfo {title} {{GNU Scientific Library Reference Manual}},}\ }
  (\bibinfo {year} {2018})\BibitemShut {NoStop}%
\bibitem [{\citenamefont {Bondi}\ \emph {et~al.}(2021)\citenamefont {Bondi}
  \emph {et~al.}}]{BONDI2021108129}%
  \BibitemOpen
  \bibfield  {author} {\bibinfo {author} {\bibfnamefont {M.}~\bibnamefont
  {Bondi}} \emph {et~al.},\ }\href {\doibase
  https://doi.org/10.1016/j.cpc.2021.108129} {\bibfield  {journal} {\bibinfo
  {journal} {Computer Physics Communications}\ }\textbf {\bibinfo {volume}
  {269}},\ \bibinfo {pages} {108129} (\bibinfo {year} {2021})}\BibitemShut
  {NoStop}%
\bibitem [{\citenamefont {Kirpichnikov}\ \emph {et~al.}(2021)\citenamefont
  {Kirpichnikov}, \citenamefont {Sieber}, \citenamefont {Molina~Bueno},
  \citenamefont {Crivelli},\ and\ \citenamefont
  {Kirsanov}}]{PhysRevD.104.076012}%
  \BibitemOpen
  \bibfield  {author} {\bibinfo {author} {\bibfnamefont {D.~V.}\ \bibnamefont
  {Kirpichnikov}}, \bibinfo {author} {\bibfnamefont {H.}~\bibnamefont
  {Sieber}}, \bibinfo {author} {\bibfnamefont {L.}~\bibnamefont
  {Molina~Bueno}}, \bibinfo {author} {\bibfnamefont {P.}~\bibnamefont
  {Crivelli}}, \ and\ \bibinfo {author} {\bibfnamefont {M.~M.}\ \bibnamefont
  {Kirsanov}},\ }\href {\doibase 10.1103/PhysRevD.104.076012} {\bibfield
  {journal} {\bibinfo  {journal} {Phys. Rev. D}\ }\textbf {\bibinfo {volume}
  {104}},\ \bibinfo {pages} {076012} (\bibinfo {year} {2021})}\BibitemShut
  {NoStop}%
\bibitem [{\citenamefont {Andreopoulos}\ \emph {et~al.}(2010)\citenamefont
  {Andreopoulos} \emph {et~al.}}]{GENIE}%
  \BibitemOpen
  \bibfield  {author} {\bibinfo {author} {\bibfnamefont {C.}~\bibnamefont
  {Andreopoulos}} \emph {et~al.},\ }\href {\doibase 10.1016/j.nima.2009.12.009}
  {\bibfield  {journal} {\bibinfo  {journal} {Nucl. Instrum. Meth. A}\ }\textbf
  {\bibinfo {volume} {614}},\ \bibinfo {pages} {87} (\bibinfo {year} {2010})},\
  \Eprint {http://arxiv.org/abs/0905.2517} {arXiv:0905.2517 [hep-ph]}
  \BibitemShut {NoStop}%
\bibitem [{\citenamefont {Brdar}\ \emph {et~al.}(2023)\citenamefont {Brdar},
  \citenamefont {Dutta}, \citenamefont {Jang}, \citenamefont {Kim},
  \citenamefont {Shoemaker}, \citenamefont {Tabrizi}, \citenamefont
  {Thompson},\ and\ \citenamefont {Yu}}]{PhysRevD.107.055043}%
  \BibitemOpen
  \bibfield  {author} {\bibinfo {author} {\bibfnamefont {V.}~\bibnamefont
  {Brdar}}, \bibinfo {author} {\bibfnamefont {B.}~\bibnamefont {Dutta}},
  \bibinfo {author} {\bibfnamefont {W.}~\bibnamefont {Jang}}, \bibinfo {author}
  {\bibfnamefont {D.}~\bibnamefont {Kim}}, \bibinfo {author} {\bibfnamefont
  {I.~M.}\ \bibnamefont {Shoemaker}}, \bibinfo {author} {\bibfnamefont
  {Z.}~\bibnamefont {Tabrizi}}, \bibinfo {author} {\bibfnamefont
  {A.}~\bibnamefont {Thompson}}, \ and\ \bibinfo {author} {\bibfnamefont
  {J.}~\bibnamefont {Yu}},\ }\href {\doibase 10.1103/PhysRevD.107.055043}
  {\bibfield  {journal} {\bibinfo  {journal} {Phys. Rev. D}\ }\textbf {\bibinfo
  {volume} {107}},\ \bibinfo {pages} {055043} (\bibinfo {year}
  {2023})}\BibitemShut {NoStop}%
\bibitem [{Ahd(2020)}]{Ahdida_2020}%
  \BibitemOpen
  \href {\doibase 10.1140/epjc/s10052-020-7788-y} {\bibfield  {journal}
  {\bibinfo  {journal} {Eur. Phys. J. C}\ }\textbf {\bibinfo {volume} {80}},\
  \bibinfo {pages} {284} (\bibinfo {year} {2020})}\BibitemShut {NoStop}%
\bibitem [{\citenamefont {Kahn}\ \emph {et~al.}(2018)\citenamefont {Kahn},
  \citenamefont {Krnjaic}, \citenamefont {Tran},\ and\ \citenamefont
  {Whitbeck}}]{Kahn:2018cqs}%
  \BibitemOpen
  \bibfield  {author} {\bibinfo {author} {\bibfnamefont {Y.}~\bibnamefont
  {Kahn}}, \bibinfo {author} {\bibfnamefont {G.}~\bibnamefont {Krnjaic}},
  \bibinfo {author} {\bibfnamefont {N.}~\bibnamefont {Tran}}, \ and\ \bibinfo
  {author} {\bibfnamefont {A.}~\bibnamefont {Whitbeck}},\ }\href {\doibase
  10.1007/JHEP09(2018)153} {\bibfield  {journal} {\bibinfo  {journal} {JHEP}\
  }\textbf {\bibinfo {volume} {09}},\ \bibinfo {pages} {153} (\bibinfo {year}
  {2018})},\ \Eprint {http://arxiv.org/abs/1804.03144} {arXiv:1804.03144
  [hep-ph]} \BibitemShut {NoStop}%
\bibitem [{\citenamefont {Alviggi}\ \emph {et~al.}(2022)\citenamefont {Alviggi}
  \emph {et~al.}}]{ShadowsLOI}%
  \BibitemOpen
  \bibfield  {author} {\bibinfo {author} {\bibfnamefont {M.}~\bibnamefont
  {Alviggi}} \emph {et~al.},\ }\href@noop {} {\emph {\bibinfo {title} {SHADOWS
  Search for Hidden And Dark Objects With the SPS}}},\ \bibinfo {type}
  {CERN-SPSC-2022-030/SPSC-I-256}\ (\bibinfo {year} {2022})\BibitemShut
  {NoStop}%
\bibitem [{\citenamefont {Cortina~Gil}\ \emph {et~al.}(2022)\citenamefont
  {Cortina~Gil} \emph {et~al.}}]{HIKE:2022qra}%
  \BibitemOpen
  \bibfield  {author} {\bibinfo {author} {\bibfnamefont {E.}~\bibnamefont
  {Cortina~Gil}} \emph {et~al.} (\bibinfo {collaboration} {HIKE}),\ }\href@noop
  {} {\  (\bibinfo {year} {2022})},\ \Eprint {http://arxiv.org/abs/2211.16586}
  {arXiv:2211.16586 [hep-ex]} \BibitemShut {NoStop}%
\bibitem [{\citenamefont {Galon}\ \emph {et~al.}(2020)\citenamefont {Galon},
  \citenamefont {Kajamovitz}, \citenamefont {Shih}, \citenamefont {Soreq},\
  and\ \citenamefont {Tarem}}]{Galon:2019owl}%
  \BibitemOpen
  \bibfield  {author} {\bibinfo {author} {\bibfnamefont {I.}~\bibnamefont
  {Galon}}, \bibinfo {author} {\bibfnamefont {E.}~\bibnamefont {Kajamovitz}},
  \bibinfo {author} {\bibfnamefont {D.}~\bibnamefont {Shih}}, \bibinfo {author}
  {\bibfnamefont {Y.}~\bibnamefont {Soreq}}, \ and\ \bibinfo {author}
  {\bibfnamefont {S.}~\bibnamefont {Tarem}},\ }\href {\doibase
  10.1103/PhysRevD.101.011701} {\bibfield  {journal} {\bibinfo  {journal}
  {Phys. Rev. D}\ }\textbf {\bibinfo {volume} {101}},\ \bibinfo {pages}
  {011701(R)} (\bibinfo {year} {2020})},\ \Eprint
  {http://arxiv.org/abs/1906.09272} {arXiv:1906.09272 [hep-ph]} \BibitemShut
  {NoStop}%
\bibitem [{\citenamefont {{Antel}}\ \emph {et~al.}(2023)\citenamefont {{Antel}}
  \emph {et~al.}}]{2023arXiv230501715A}%
  \BibitemOpen
  \bibfield  {author} {\bibinfo {author} {\bibfnamefont {C.}~\bibnamefont
  {{Antel}}} \emph {et~al.},\ }\href {\doibase 10.48550/arXiv.2305.01715}
  {\bibfield  {journal} {\bibinfo  {journal} {arXiv e-prints}\ ,\ \bibinfo
  {eid} {arXiv:2305.01715}} (\bibinfo {year} {2023})},\ \Eprint
  {http://arxiv.org/abs/2305.01715} {arXiv:2305.01715 [hep-ph]} \BibitemShut
  {NoStop}%
\end{thebibliography}%

\end{document}